\documentstyle[preprint,prd,aps,tighten,epsf]{revtex}
\begin{document}
\draft
\newcommand{\nl}{\nonumber \\}
\newcommand{\bea}{\begin{eqnarray}}
\newcommand{\eea}{\end{eqnarray}}
\newcommand{\bi}{\bibitem}
\newcommand{\be}{\begin{equation}}
\newcommand{\ee}{\end{equation}}
\newcommand{\bt}{\begin{table}}
\newcommand{\et}{\end{table}}
\newcommand{\btab}{\begin{tabular}}
\newcommand{\etab}{\end{tabular}}
\def\eg{{\it e.g.}}
\def\ra{\rightarrow}
\def\dek#1{\times10^{#1}}
\def\Vus{\left|V_{us}\right|}
\def\ks{K^0_S}
\def\kl{K^0_L}
\def\ksp{K^{*+}}
\def\ksm{K^{*-}}
\def\ksmf{K^*(892)^-}
\def\knsm{K_0^{*-}}
\def\knsmf{K_0^*(1430)^-}
\def\ts{\tau^-\ra\knsm\nu_\tau}
\def\kml3full{K^-\ra \pi^0 \ell^- \bar{\nu}_\ell}
\def\knl3full{\bar{K}^0\ra \pi^+ \ell^- \bar{\nu}_\ell}
\def\kl3{K_{\ell3}}
\def\kcm3{K^\pm_{\mu3}}
\def\klm3{K^0_{\mu3}}
\def\kce3{K^\pm_{e3}}
\def\kle3{K^0_{e3}}
\def\tauck{\tau^-\ra K^-\pi^0\nu_\tau}
\def\taunk{\tau^-\ra \bar{K}^0\pi^-\nu_\tau}

\def \zpc#1#2#3{Z.~Phys.~C {\bf#1}, #2 (19#3)}
\def \plb#1#2#3{Phys.~Lett.~B {\bf#1}, #2 (19#3)}
\def \prl#1#2#3{Phys.~Rev.~Lett.~{\bf #1}, #2 (19#3)}
\def \pr#1#2#3{Phys.~Rep.~{\bf #1}, #2 (19#3)}
\def \prd#1#2#3{Phys.~Rev.~D~{\bf#1}, #2 (19#3)}
\def \npb#1#2#3{Nucl.~Phys.~B {\bf#1}, #2 (19#3)}
\def \rmp#1#2#3{Rev.~Mod.~Phys.~{\bf#1}, #2 (19#3)}
\def \ea{{\it et al.}}
\def \ibid#1#2#3{{\it ibid.} {\bf#1}, #2 (19#3)}
\def \sjnp#1#2#3#4{Yad. Fiz. {\bf#1}, #2 (19#3) [Sov. J. Nucl. Phys.
{\bf#1}, #4 (19#3)]}
\def \pan#1#2#3#4{Yad. Fiz. {\bf#1}, #2 (19#3) [Physics of Atomic Nuclei
{\bf#1}, #4 (19#3)]}
\def \epjc#1#2#3{Eur. Phys. J. C {\bf#1}, #2 (19#3)}
\def \appb#1#2#3{Acta Phys. Pol. B {\bf#1}, #2 (19#3)}

\title{\bf 
Relations between the $K_{\ell 3}$ and $\tau\ra K\pi\nu_\tau$ decays
}
\author{Peter Lichard\footnote{On leave of absence from Department of 
Theoretical Physics, Comenius University,
842-15 Bratislava, Slovak Republic.}}
\address{
Department of Physics and Astronomy, University of Pittsburgh,
Pittsburgh, Pennsylvania 15260 \\
and Institute of Physics, Silesian University, 746-01 Opava, Czech Republic
}
\maketitle
\begin{abstract}
We investigate the relations between the $\kl3$ and $\tau\ra K\pi\nu_\tau$ 
decays using the meson dominance approach. First, the
experimental branching fractions (BF) for $\kce3$ and $\kle3$ are used 
to fix two normalization constants (isospin invariance is not
assumed). Then, the BF of $\tau^-\ra
\ksmf\nu_\tau$ is calculated in agreement with experiment. We further
argue that the nonzero value of the slope parameter $\lambda_0$ of the
$\kcm3$ and $\klm3$ form factors $f_0(t)$ implies the existence
of the $\tau^-\ra \knsmf \nu_\tau$ decay. We calculate its BF, together
with BF's of the $\kcm3$, $\klm3$, $\tauck$, and $\taunk$ decays, as
a function of the $\lambda_0$ parameter. At some value of $\lambda_0$,
different for charged and neutral kaons, calculated BF's seem to match
existing data and a prediction is obtained for the $\tau\ra K\pi\nu$
decays going through the $\knsmf$ resonance.

\end{abstract}


\narrowtext

With a new generation of high statistics and precise data about the
$\kl3$, {\it i.e.} $K\ra\pi\ell\nu_\ell$, decays coming soon 
\cite{dafne,e865} 
it is possible to think about investigating the problems that were not 
fully resolved in the previous series of experiments, which ended
approximately in the early eighties. 

One of the as yet undecided issues is that of the value, or even the sign,
of the slope $\lambda_0$ in the linear parametrization of the form factor 
$f_0$, the definition of which we give below. Some $\kcm3$ experiments 
indicated a non-vanishing negative value, some positive.\footnote{We refer
the reader to \cite{pdg} for references and more details.} The situation was 
analysed by the Particle Data Group in 1982 \cite{pdg1982} and a recommended 
value of $0.004\pm0.007$ was chosen. A very recent experiment \cite{artemov}
with its result of $0.062\pm0.024$ influenced the recommended value, which 
has now become $0.006\pm0.007$ \cite{pdg}. 

The situation with the $\lambda_0$ parameter in the 
$K_L^0\ra\pi^\pm\mu^\mp\nu$ ($\klm3$) decay seems to be 
a little more definite, at least judging from the recommended value of
$0.025\pm0.006$ \cite{pdg1982,pdg} and from all the experiments in the period
of 1974-1981 agreeing on the positive sign. 

In this note we speculate about consequences which may stem from conclusive 
establishing a nonzero value of $\lambda_0$. Its purpose is not to
compete with the elaborate calculations of the $K_{\ell3}$ form factors,
see \cite{gassleut,shabalin,bijnens,buck}, or of the kaon production
in $\tau$-lepton decays \cite{tauintokaons,scalar-to-tau}. Our aim is to 
show on
a phenomenological basis in a simple and transparent way possible relations
between the $\kl3$ and $\tau\ra K\pi\nu_\tau$ decays. We mainly argue 
that a nonzero value of the $\lambda_0$ parameter of the $K_{\mu3}$ decays 
implies a nonzero decay fraction of the $\ts$ decay. Judging from our results
and the contemporary experimental upper limit, this decay may be observed
soon. The tool we are going to use here is 
the meson dominance hypothesis, see \cite{md} and references therein.

If we believe in the validity of the standard electroweak model in the 
leptonic sector, we parametrize the matrix element of the 
$\kl3$ decay in the form \cite{macdowell57,macdowell59} 
\be
\label{mekl3}
{\cal M}_{K_{\ell3}}= C \left[f_+(t)\ p^\mu+ f_-(t)\ q^\mu\right]\ 
\bar{u}\gamma_\mu(1-\gamma_5)v\ ,
\ee
where $p$ ($q$) is the sum (difference) of the four-momenta of the $K$ and 
$\pi$ mesons, $t=q^2$, and $u$ and $v$ are appropriately chosen Dirac spinors 
of outgoing leptons. This relation defines, up to a normalization factor,
the $\kl3$ form factors $f_+(t)$ and $f_-(t)$. The normalization
used most frequently \cite{leutroos,bijnens}
is defined by $C=G_F|V_{us}|/2$ for the $K^\pm_{\ell3}$
and $C=G_F|V_{us}|/\sqrt{2}$ for the $K^0_{\ell3}$ decays.
It is customary to introduce also the form factor \cite{macdowell59}
\be
f_0(t)=f_+(t)+\frac{t}{m_K^2-m_\pi^2}\ f_-(t)\ ,
\ee
which corresponds to the $J=0$ state of the $K-\pi$ system, whereas
$f_+(t)$ to its $J=1$ state. After integrating over
angular variables, the differential decay rate in $t$, which has also
a meaning of the invariant mass squared of the $\ell\nu$ system,
comes out as
\bea
\label{dgamdt1a}
\frac{d\Gamma_{\kl3}}{dt}&=&\frac{C^2}{3(4\pi m_K)^3}
\frac{\left(t-m_\ell^2\right)^2}{t^3}\  
\lambda^{1/2}(t,m_K^2,m_\pi^2) \\
&\times&
\left[\left(2t+m_\ell^2\right)\lambda(t,m_K^2,m_\pi^2)\left|f_+(t)\right|^2
+3m_\ell^2\left(m_K^2-m_\pi^2\right)^2\left|f_0(t)\right|^2\right]
\ ,\nonumber
\eea
where $\lambda(x,y,z)=x^2+y^2+z^2-2xy-2xz-2yz$.
The $t$-dependence of all form factors is usually studied 
experimentally in linear approximation 
\be
\label{linear}
f(t)=f(0)\left(1+\lambda\ \frac{t}{m_\pi^2}\right)\ ,
\ee
although such an approximation was shown \cite{chounet} to be improper, 
at least for the $f_+(t)$ form factor of the $\kce3$ and $\kle3$ decays. 
The authors of \cite{chounet} found big discrepancies among $\lambda_+$'s 
from different experiments if a linear approximation was used. They clearly 
demonstrated the existence of a quadratic term in $f_+(t)$ by showing
that its inclusion led to better fits.
 
There is a peculiarity of the present experimental situation, which
is worth mentioning. The $\mu/e$ universality requires 
the form factors be equal for the $K_{e3}$ and $K_{\mu3}$ decays.
Assuming the validity of (\ref{linear}) we can express the 
$R=K_{\mu3}/K_{e3}$ branching ratio as a function of two parameters:
$\lambda_+$ and $\lambda_0$. Knowing the experimental values of the latter
we can evaluate $R$ and compare it with the experimental ratio.
The $K^\pm_{\ell3}$ data pass this consistency check without problems, 
whereas the contemporary recommended values of the $K^0_{\ell3}$ form 
factor slopes lead to a little lower ratio than the experimental one
($0.676\pm0.009$ against $0.701\pm0.008$). To restore the consistency,
one has to sacrifice the $\mu/e$ universality and allow a higher value 
of the $\lambda_+$ parameter in the $\klm3$ decay.

A remark is required at the very beginning about our treatment of the 
$\kl3$ decays of neutral
kaons. We will work with the $K^0\ra\pi^-\ell^+\nu_\ell$
and $\bar{K}^0\ra\pi^+\ell^-\bar{\nu}_\ell$ decays, despite the fact
that what is really observed are decays of the $K^0_L$ and $K^0_S$ mesons.
If we ignore a small violation of the $CP$ invariance, then the
decay rates of the former two decays are identical and each of them
is equal to the decay rate of $K_L^0\ra\pi^\pm\ell^\mp\nu$, where summing 
is understood over the two final states shown. The same is true for
$K_S^0\ra\pi^\pm\ell^\mp\nu$.

The assumption that the $\kl3$ decay is dominated by the $K^*(892)$ pole,
pictorially depicted in Fig.~1a, leads to the following matrix element
(see, e.g., \cite{kstardom,md}):
\be
\label{mekl31a}
{\cal M}_{1a}=\frac{G_{a}}{m_V^2-t}\left(p^\mu-
\frac{m_K^2-m_\pi^2}{m_V^2}\ q^\mu\right)
\bar{u}\gamma_\mu(1-\gamma_5)v\ ,
\ee
where $m_V$ is the $K^{*\pm}(892)$ mass and (dimensionless) $G_a$ collects 
the coupling constants from all vertices. It includes also the $V_{us}$ 
element of the
Cabibbo-Kobayashi-Maskawa matrix. As the isospin invariance is
badly broken in the $\kl3$ decays, see, \eg, the discussion in \cite{leutroos},
we have two independent constants. One for $\kl3$ decays of $K^\pm$, another 
for $K^0$ ($\bar{K}^0$). We do not need the explicit form of $G_a$'s, because 
we will fix their values from the experimental values of the corresponding 
$K_{e3}$ decay rates. Nevertheless, in the notation of Ref.~\cite{md} we have
\be
G_a^{(\pm)}=G_FV_{us}w_{K^*}m_V^2\frac{g_{K^{*\pm}K^\pm\pi^0}}{g_\rho} \ 
\ee
and a similar relation for $G_a^{(0)}$. The connection with the standard
notation \cite{leutroos,bijnens} is given by $G_a^{(\pm)}/m_V^2=
G_F\Vus f_+^{K^+\pi^0}(0)/2$. For $G_a^{(0)}$, the factor of 2 is replaced
by $\sqrt{2}$. 

Let us note that when writing (\ref{mekl31a}) we took the propagator
of the $K^*$ resonance in the free-vector-particle form 
\be
\label{freepropagator}
-iG^{\mu\nu}_0(q)=\frac{-g^{\mu\nu}+q^\mu q^\nu/m_V^2}{t-m_V^2+i\epsilon}\ ,
\ee
where $m_V$ is the mass of the $\ksmf$ resonance, as seen in the 
hadronic production experiments.
The absence of a noninfinitesimal imaginary part in denominator 
is justified by $t$ being below the threshold of the $K^*\ra K\pi$ decay
channel. But the actual form of the propagator may differ from 
(\ref{freepropagator}) even in the subtreshold region.
The success in describing the $\kl3$ form factors gives an 
{\it a posteriori} phenomenological argument in favor of 
an approximate validity of Eq. (\ref{freepropagator}).

If we fix, for simplicity, the normalization of the form factors by 
requiring $f_+(0)=1$, we find the following correspondence of 
(\ref{mekl31a}) with the quantities entering Eq. (\ref{mekl3}): 
\bea
\label{ff1a}
C&=&\frac{G_a}{m_V^2}\ ,\\
f_+(t)&=&\frac{m_V^2}{m_V^2-t}\ , \nl
f_-(t)&=&-\frac{m_K^2-m_\pi^2}{m_V^2-t}\ . \nonumber
\eea
We also have
\be
\label{f0eq1}
f_0(t)=1\ . 
\ee
Inserting our $C$, $f_+(t)$, and $f_0(t)$ to the general formula
(\ref{dgamdt1a}), integrating over $t$, and comparing our result
with the $\kce3$ ($\kle3$) decay rate calculated from the experimental
values of the $K^\pm$ ($K_L^0$) lifetime and the $\kce3$ ($\kle3$)
branching fraction we arrive at 
${G_a^{(\pm)}}^2=(1.037\pm0.013)\dek{-12}$ and 
${G_a^{(0)}}^2=(1.974\pm0.021)\dek{-12}$. If the isospin invariance
in the $K^*K\pi$ vertex were exact, the ratio of the
former to the latter would be equal to 1/2. 

Before proceeding further with our form-factor issue let us notice
that the same overall coupling constants govern also the
decays $\tauck$ and $\taunk$ in which the $K\pi$ system
is produced via the $\ksm$ resonance, see Fig. 2a. Let us first
calculate their branching fractions using the $G_a$'s  we have just 
determined. This will test the soundness of our approach and of 
the approximations made and will give us the confidence for calculations
for which the comparison with data is impossible as yet.

The main problem we are faced with when attempting such a calculation
is that of the propagators of resonances. We are now above the threshold 
of the $K\pi$ system, $s>(m_K+m_\pi)^2$, where $s$ is the square of
the four-momentum $p$ flowing through the $K^*$ resonance. As a consequence, 
the propagator acquires an important imaginary part and may differ 
substantially from the propagator of a free vector particle also in 
other respects. For example, in \cite{wzprop} it was proposed that 
the lowest order $W^\pm$ ($Z^0$) renormalized propagator in the 
unitary gauge can be obtained, at least in the resonance region,
by a simple modification of the free propagator (\ref{freepropagator}).
Namely, by replacing the mass squared $m_V^2$ 
everywhere in Eq.~(\ref{freepropagator}) by $m_V^2-im_V\Gamma_V$,
with $\Gamma_V$ being the resonance width.\footnote{For later
development and references to alternative approaches to the
weak-gauge-bosons propagators see Ref.~\cite{wprop}.}
For resonances with strong interaction such a simple prescription
is not justified, as discussed, \eg, in \cite{isgurtau}. Nevertheless,
if $s=p^2$ is in a close proximity to the resonant mass squared we can 
write
\be
\label{vresprop}
-iG^{\mu\nu}(p)=\frac{-g^{\mu\nu}+\omega(s)p^\mu p^\nu/s}
{s-m_V^2+im_V\Gamma_V(s)}\ ,
\ee
where $\Gamma_V(s)$ is the $s$-dependent total width of the resonance
normalized by $\Gamma(m_V^2)=\Gamma_V$ and $\omega(s)$ is a complex 
function. It reflects the
properties of the one-particle-irreducible bubble and is, in
principle, calculable. There are different ways of treating it 
in practice. For example, when considering the $a_1$ resonance in
the intermediate state, the authors of \cite{isgurtau} eliminated
its influence by choosing transverse vertices. Alternatively,
various choices have been made in the literature. Very popular
is the free-particle choice $\omega(s)=s/m_V^2$, recently used,
\eg, in Ref.~\cite{scalar-to-tau}. In experimental analyses
a spin-zero propagator is used even where not justified
(see discussion in \cite{aps}). This corresponds to $\omega(s)=0$. 
The same choice was made in \cite{nava}, where the branching
fraction of the $\tau^-\ra \ksmf\nu_\tau$ decay was also calculated.

Fortunately, the $K^*(892)$ resonance is relatively narrow 
($\Gamma_V\approx 51$~MeV) and we can hope that the systematic error
connected with the propagator ambiguity is small. Nevertheless,
to assess it we will calculate every quantity of interest twice.
Once with $\omega=s/[m_V^2-im_V\Gamma_V(s)]$, then with $\omega=0$. 
This procedure yields an average and an estimate of its systematic error.

The differential rate of the $\tauck$ decay in the mass squared of the 
$K\pi$ system is given by the formula
\bea
\label{dgamds2}
\frac{d\Gamma_{\tauck}}{ds}&=&\frac{1}{6(4\pi m_\tau)^3}
\frac{\left(m_\tau^2 -s\right)^2}{s^3}\ \lambda^{1/2}(s,m_K^2,m_\pi^2)\\
&\times&
\left[\left(2s+m_\tau^2\right)\lambda(s,m_K^2,m_\pi^2)\left|F_+(s)\right|^2
+3m_\tau^2\left(m_K^2-m_\pi^2\right)^2\left|F_0(s)\right|
^2\right]\ ,\nonumber
\eea
where
\be
\label{f1}
F_+(s)=\frac{G_a^{(\pm)}}{s-m_V^2+im_V\Gamma_V(s)}
\ee
and
\be
\label{f2a}
F_0(s)=\frac{G_a^{(\pm)}\left[1-\omega(s)\right]}
{s-m_V^2+im_V\Gamma_V(s)}\ .
\ee
The presence of $F_0(s)$ in (\ref{dgamds2}) reflects the contribution
of the off-mass-shell vector resonance $K^*$ to the $J=0$ channel.
It would disappear if we chose $\omega(s)\equiv 1$, as seen from 
(\ref{f2a}).
After integrating (\ref{dgamds2}) and using the experimental value 
of the $\tau^-$ lifetime, we arrive at $B(\tauck)=(3.9\pm0.6)\dek{-3}$. 
We proceed similarly to obtain $B(\taunk)=(7.1\pm1.2)\dek{-3}$. 
After summing these two branching fractions we get
\be
B(\tau^-\ra \ksmf\nu_\tau)=(1.10\pm0.18)\%\ .
\ee
The experimental value \cite{pdg} is $(1.28\pm0.08)\%$. 

Let us now return to the form factors.
The salient feature of the one-vector-meson dominance model
is the constant $\kl3$ form factor $f_0$, which implies a vanishing
parameter $\lambda_0$ defined in (\ref{linear}). There are at least 
two ways to accommodate a nonvanishing value of $\lambda_0$ in the
meson dominance approach. 

One possibility is to add more strange vector
resonances. The case of two vector resonances was considered already
in \cite{kstardom}. In addition to the well established $K^*(892)$
it was $K^*(730)$, which was abandoned later on. But the formulas
of \cite{kstardom} are general, and could be used for inclusion
of $K^*(1410)$ as well. 

Another way of modifying the meson dominance approach to the
$\kl3$ decay is to include the scalar resonance $K^*_0(1430)$.
The advantage of this approach is that, as we will see, it
does not modify the $f_+(t)$ form factor, which seems to be well
described already with the $K^*(892)$ alone. The modification
influences only the $f_-(t)$ and, consequently, the $f_0(t)$ 
form factors. Already the authors of \cite{jth} discussed this
possibility, but at that time there was no known $K$-$\pi$ resonance
with spin zero. 

To calculate the contribution to the $\kl3$ matrix element
from the Feynman diagram with the $\knsmf$ in the intermediate
state (Fig. 1b), let us first define the weak decay constant of the
$K_0^{*-}$. As usual, it can be done by means of the matrix
element of the vector part of the strangeness-changing quark current
\be
\label{fk0st}
\langle 0|\bar{u}(0)\gamma^\mu s(0)|p\rangle_{K_0^{*-}}=
i\ f_{K_0^{*-}}\ p^\mu\ .
\ee
Then, the diagram in Fig.~1b yields
\be
\label{mekl31b}
{\cal M}_{1b}=\frac{G_b}{m_S^2-t}\ q^\mu
\bar{u}\gamma_\mu(1-\gamma_5)v\ ,
\ee
where $m_S$ is the $\knsmf$ mass and 
\be
G_b=\frac{G_F}{\sqrt 2}V_{us}f_{K_0^{*-}}
g_{K_0^{*-}K^-\pi^0}\ .
\ee
Because (\ref{mekl31b}) does not contain $P^\mu$, the constant $C$ and 
the form factor $f_+(t)$, as shown in (\ref{ff1a}), will not change after 
adding (\ref{mekl31b}) to (\ref{mekl31a}). New $f_-(t)$ and $f_0(t)$
become 
\bea
f_-(t)&=&-\frac{m_K^2-m_\pi^2}{m_V^2-t}\ +\frac{G_b}{G_a}
\frac{m_V^2}{m_S^2-t}\ ,\\
f_0(t)&=&1+\frac{G_b}{G_a}\frac{m_V^2}{m_K^2-m_\pi^2}\frac{t}{m_S^2-t}
\nonumber \ .
\eea
The parameter $\lambda_0$ now acquires the value
\be
\lambda_0=\frac{G_b}{G_a}\frac{m_V^2}{m_S^2}
\frac{m_{\pi}^2}{(m_K^2-m_\pi^2)} \ .
\ee
We see that the nonzero weak decay constant of $\knsm$ leads to deviation
of the $\lambda_0$ parameter from zero. But to check whether a nonvanishing
value of $\lambda_0$ is really caused by a $\knsm$ in the intermediate
state of the $\kl3$ decay, we must look for other consequences of the
weak interaction of $\knsm$ and their consistency with the $\kl3$ decay
phenomenology. The most obvious candidate for such a program is the
decay of $\tau^-$ lepton to neutrino and $\knsm$. Or, to be more precise,
to the $K^-\pi^0$ system which originates from the strong decay of 
$\knsm$. 

When calculating the branching fraction of the $\tauck$ and $\taunk$ decays, 
we should include the possible interference between the $\ksmf$ 
and $\knsmf$ channels, {\it i.e.}, add coherently the diagrams (a) and 
(b) shown in Fig.~\ref{taufig}. The resulting differential decay rate formula 
for $\tauck$ coincides with Eq. (\ref{dgamds2}). Function $F_+(s)$ is again 
given by (\ref{f2a}) because the scalar resonance cannot contribute to 
the $J=1$ channel, but
\be
\label{f2ab}
F_0(s)=\frac{G_a^{(\pm)}\left[1-\omega(s)\right]}{s-m_V^2+im_V\Gamma_V(s)}
+\frac{G_b^{(\pm)}}{m_K^2-m_\pi^2}\frac{s}{s-m_S^2+im_S\Gamma_S(s)}\ .
\ee
The changes needed to get a formula for the same quantity in $\taunk$
are obvious.

Now we have all necessary formulas and constants prepared and can calculate
the quantities of interest for various values of the slope parameter
$\lambda_0$. The results are shown in Tab.~\ref{tab1} for the charged
kaons, in Tab.~\ref{tab2} for the neutral kaons.

Inspecting Tab.~\ref{tab1} we see that to get simultaneously the correct 
branching fraction of both $\kcm3$ and $\tauck$ decays, we need to pick
$\lambda_0\approx$~0.020. This is higher than the present recommended
value $(6\pm7)\dek{-3}$. But with eyes on the recent experiment 
\cite{artemov} with its $0.062\pm0.024$, we do not consider disastrous the 
dicrepancy of our value of $\lambda_0$ with the recommended one. 
Our value also agrees with $\lambda_0=0.019$ obtained
on the basis of the Callan-Treiman relation \cite{ct}, see \cite{bijnens}. 
With reference to the experiment \cite{artemov} it should be said that
$\lambda > 0.04$ contradicts the estimate of the upper 
limit for the non-$\ksmf$ $K^-\pi^0$ production in $\tau^-$ decays. 
On the basis of $\lambda_0\approx0.020$ we expect the branching fraction 
for producing the $K^-\pi^0$ system in $\tau^-$ decays via the scalar 
$\knsmf$ resonance to be $\approx2\dek{-4}$.

Similar analysis of numbers in Tab.~\ref{tab2} points to a $\lambda_0$
for the $\klm3$ decay somewhere around 0.030, which is in agreement
with the recommended value \cite{pdg}, but higher than in the previous
case. The higher value is required by the $\klm3$ branching fraction. 
As a consequence, also the branching fraction of the $\bar{K}^0\pi^-$
production from the $\tau^-\ra\knsmf\nu_\tau$ decay, $\approx8\dek{-4}$,
is higher than would correspond to $K^-\pi^0$ and isospin symmetry.

On the basis of our estimates we expect the branching fraction
of the $\tau^-\ra\knsmf\nu_\tau$ decay to be around 0.1\%.

In Fig.~\ref{massspectrum} we show the mass spectrum of the $\bar{K}^0\pi^-$
system produced in the $\taunk$ decays assuming $\lambda_0=0.030$.
We concentrate on the $\knsmf$ mass region to show different 
contributions to the final yield. The tail of the $\ksmf$ resonance
modifies the resonance shape significantly, whereas the interference
between the two contributing intermediate states is negligible.

We hope that in the near future the high statistics and precise kaon 
decay data on one side, and data from the $\tau$-factories 
\cite{taufactory}
on the other, will enable to study the relations between the 
$K_{\ell3}$ and $\tau\ra K\pi\nu$ decays in more detail.
 
Finally, we would like to comment the role of the meson dominance
model. It is clear that this approach cannot substitute for a more
fundamental theory based on first principles. We cannot even say
in advance for which processes it will offer a fair
description and for which it will fail. But, in our opinion, it has
an important role as a heuristic tool. In the cases where it
succeeds, it shows which underlying quark diagrams are most important
for understanding the dynamics of the process. Because of the necessity
of convoluting the simple pure electroweak diagrams with QCD dynamics
in order to form hadrons, the diagrams which seem to be important
on the quark level, may finally become unimportant and vice versa.
We discussed this aspect in some detail in \cite{md} in connection with 
the $K^+\ra\pi^+e^+e^-$ decay. Also here, the ability of the meson
dominance to describe, with the same set of basic parameters, both
$K_{\ell3}$ and $\tau^-\ra K\pi\nu$ decays hints that the most 
important mechanism of destroying strangeness is the
quark-antiquark annihilation to the $W$ boson.
This picture differs completely from the usual notion in which the 
non-strange quark is a spectator and proceeds through the process
intact, whereas the strange quark converts to a non-strange one 
by emitting $W$.

\acknowledgements
The author is indebted to Dave Kraus and Julia Thompson for discussions.
This work was supported by the U.S. Department of Energy under contract No. 
DOE/DE-FG02-91ER-40646 and by the Grant Agency of the Czech Republic under
contract No. 202/98/0095. The hospitality of the CERN Theory Division, where
a part of this work was done, is gratefully acknowledged.

\bt
\caption{Branching fractions of the $\kcm3$ and $\tauck$ decays calculated
within the meson dominance approach assuming various values of the
$\kcm3$ parameter $\lambda_0$. The  recommended experimental values 
\protect\cite{pdg} are shown in the last row.}
\label{tab1}
\btab{cccc}
$\lambda_0\dek{3}$ & $B(\kcm3)$ & 
$ B(\tau^-\ra K^-\pi^0\nu_\tau)$ &
$ B(\tau^-\ra K^-\pi^0\nu_\tau)\dek{3}$ \\
  & (\%) &  via $K_0^*(1430)^-$ & total (a)  \\
\hline
  -10&$3.03\pm0.02$&$ 5.3\dek{-5}$&$ 3.9\pm 0.6$ \\
   -5&$3.06\pm0.02$&$ 1.3\dek{-5}$&$ 3.9\pm 0.6$ \\
    0&$3.10\pm0.02$&    0         &$ 3.9\pm 0.6$ \\
    5&$3.13\pm0.02$&$ 1.3\dek{-5}$&$ 3.9\pm 0.6$ \\
   10&$3.17\pm0.02$&$ 5.3\dek{-5}$&$ 4.0\pm 0.6$ \\
   15&$3.21\pm0.02$&$ 1.2\dek{-4}$&$ 4.0\pm 0.6$ \\
   20&$3.25\pm0.02$&$ 2.1\dek{-4}$&$ 4.1\pm 0.6$ \\
   25&$3.29\pm0.02$&$ 3.3\dek{-4}$&$ 4.3\pm 0.6$ \\
   30&$3.33\pm0.02$&$ 4.8\dek{-4}$&$ 4.4\pm 0.6$ \\
   35&$3.37\pm0.02$&$ 6.5\dek{-4}$&$ 4.6\pm 0.6$ \\
   40&$3.41\pm0.02$&$ 8.4\dek{-4}$&$ 4.8\pm 0.6$ \\
   45&$3.45\pm0.02$&$ 1.1\dek{-3}$&$ 5.1\pm 0.6$ \\
   50&$3.49\pm0.02$&$ 1.3\dek{-3}$&$ 5.3\pm 0.6$ \\
   55&$3.53\pm0.02$&$ 1.6\dek{-3}$&$ 5.6\pm 0.6$ \\
   60&$3.58\pm0.02$&$ 1.9\dek{-3}$&$ 6.0\pm 0.6$ \\
\hline
$6\pm7$  &$3.18\pm0.08$ &$<9\dek{-4}$ (b)& $5.2\pm0.5$ 
\etab
(a) Total = $\ksmf$ + $\knsmf$ + interference term.\\
(b) Estimated as a half of $\tau^-\ra\pi^-\bar K^0\nu_\tau$, 
non-$K^*(892)^-$.
\et

\bt
\caption{Branching fractions of the $\klm3$ and $\taunk$ decays calculated
within the meson dominance approach assuming various values of the
$\klm3$ parameter $\lambda_0$. The  recommended experimental values 
\protect\cite{pdg} are shown in the last row.}
\label{tab2}
\btab{cccc}
$\lambda_0\dek{3}$ & $B(\klm3)$ & 
$ B(\taunk)$ &
$ B(\taunk)\dek{3}$ \\
  & (\%) &  via $K_0^*(1430)^-$ & total (a)  \\
\hline
  -10&$24.39\pm 0.17$&$ 8.9\dek{-5}$&$  7.2\pm 1.2$ \\
   -5&$24.65\pm 0.17$&$ 2.2\dek{-5}$&$  7.1\pm 1.2$ \\
    0&$24.91\pm 0.17$& 0 &$  7.1\pm 1.2$ \\
    5&$25.18\pm 0.18$&$ 2.2\dek{-5}$&$  7.2\pm 1.2$ \\
   10&$25.46\pm 0.18$&$ 8.9\dek{-5}$&$  7.3\pm 1.1$ \\
   15&$25.74\pm 0.18$&$ 2.0\dek{-4}$&$  7.4\pm 1.1$ \\
   20&$26.02\pm 0.18$&$ 3.6\dek{-4}$&$  7.6\pm 1.1$ \\
   25&$26.31\pm 0.18$&$ 5.6\dek{-4}$&$  7.8\pm 1.1$ \\
   30&$26.61\pm 0.19$&$ 8.0\dek{-4}$&$  8.1\pm 1.1$ \\
   35&$26.91\pm 0.19$&$ 1.1\dek{-3}$&$  8.4\pm 1.1$ \\
   40&$27.21\pm 0.19$&$ 1.4\dek{-3}$&$  8.8\pm 1.1$ \\
   45&$27.52\pm 0.19$&$ 1.8\dek{-3}$&$  9.2\pm 1.1$ \\
   50&$27.84\pm 0.19$&$ 2.2\dek{-3}$&$  9.6\pm 1.1$ \\
   55&$28.15\pm 0.20$&$ 2.7\dek{-3}$&$ 10.1\pm 1.1$ \\
   60&$28.48\pm 0.20$&$ 3.2\dek{-3}$&$ 10.7\pm 1.1$ \\
\hline
$25\pm6$  &$27.17\pm0.25$ & $<1.7\dek{-3}$ (b)& $8.3\pm0.8$ 
\etab
(a) Total = $\ksmf$ + $\knsmf$ + interference term.\\
(b) Non-$\ksmf$ $\nu_\tau$.
\et

\begin{figure}
\begin{center}
\leavevmode
\setlength \epsfysize{10cm}
\setlength \epsfxsize{15cm}
\epsffile{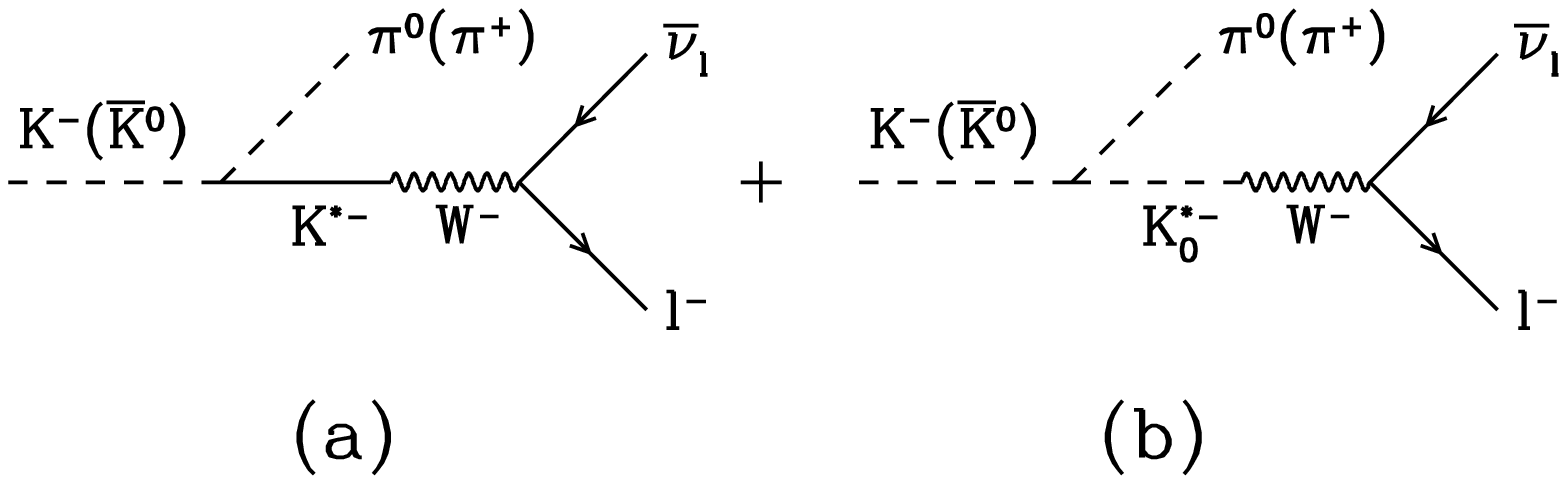}
\end{center}
\caption{Feynman diagrams contributing to the matrix element of the
\protect{$\kml3full$} ($\knl3full$) decay with (a) the vector 
resonance $\ksmf$ and (b) the scalar resonance $\knsmf$ in the
intermediate state.
}
\label{kl3fig}
\end{figure}

\begin{figure}
\begin{center}
\leavevmode
\setlength \epsfysize{10cm}
\setlength \epsfxsize{15cm}
\epsffile{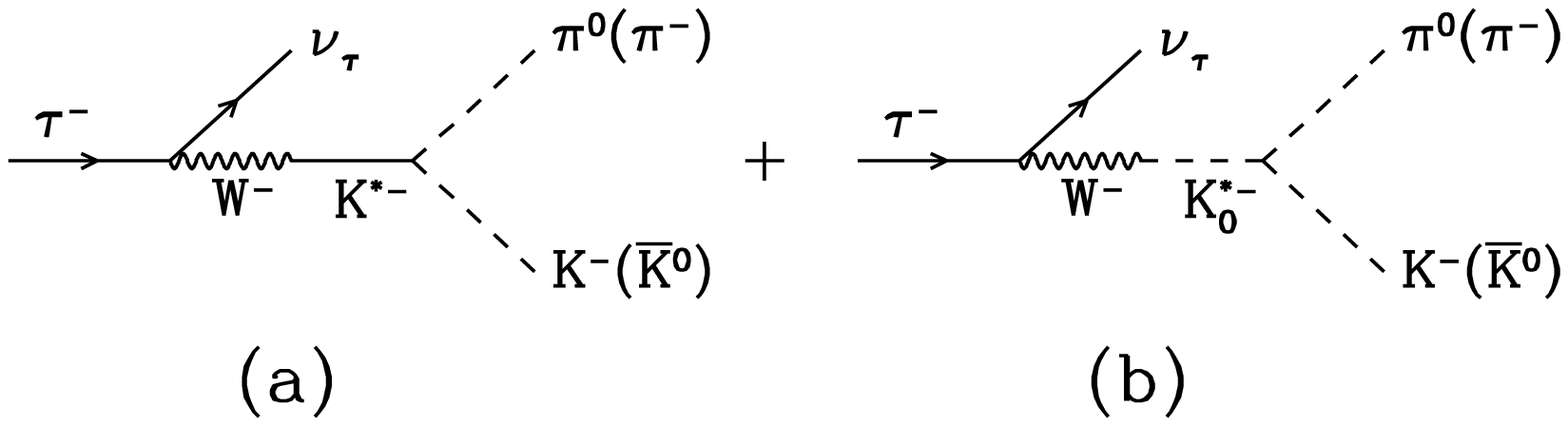}
\end{center}
\caption{Feynman diagrams contributing to the matrix element of the
\protect{$\tauck$} ($\taunk$) decay with (a) the vector 
resonance $\ksmf$ and (b) the scalar resonance $\knsmf$ in the
intermediate state.
}
\label{taufig}
\end{figure}

\begin{figure}
\begin{center}
\leavevmode
\setlength \epsfysize{15cm}
\setlength \epsfxsize{15cm}
\epsffile{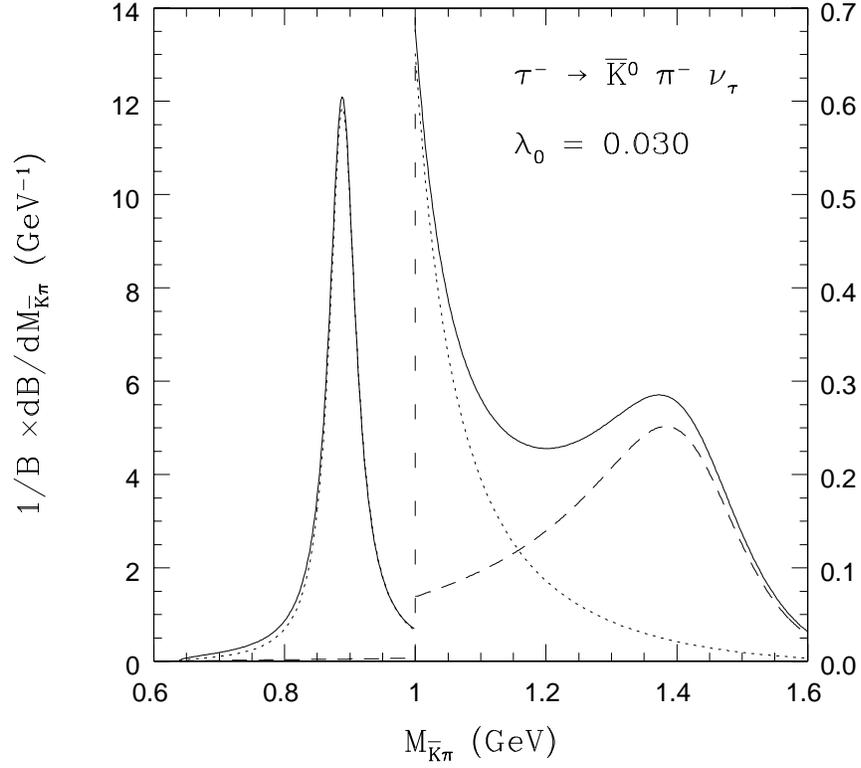}
\end{center}
\caption{Mass spectrum of the $\bar{K}\pi$ system produced in
the $\taunk$ decay with $\ksmf$ and $\knsmf$ in the intermediate 
state assuming the $\klm3$ parameter $\lambda_0=0.030$. Solid
curve: the total branching fraction; dotted curve: $\ksmf$ only; dashed 
curve: $\knsmf$ only. Notice a different scale for masses above 1 GeV.}
\label{massspectrum}
\end{figure}

\end{document}